\documentclass[10pt]{article}

\usepackage{amsmath}
\usepackage{amssymb}

\usepackage{graphicx}
\usepackage{epstopdf}
\usepackage{subfigure}

\usepackage{cite}
\usepackage{url}
\usepackage{color} 

\usepackage[labelfont=bf,labelsep=period,justification=raggedright]{caption}

\makeatletter
\renewcommand{\@biblabel}[1]{\quad#1.}
\makeatother

\usepackage{tikz}
  \usetikzlibrary{shapes,shadows}
  \tikzstyle{abstractbox} = [draw=black, fill=white, rectangle, 
  inner sep=10pt, style=rounded corners, drop shadow={fill=black,
  opacity=1}]
  \tikzstyle{abstracttitle} =[fill=white]
 
  \newcommand{\boxabstract}[2][fill=white]{
    \begin{center}
      \begin{tikzpicture}
        \node [abstractbox, #1] (box)
        {\begin{minipage}{\linewidth}
            \normalsize #2
          \end{minipage}};
        \node[abstracttitle, right=10pt] at (box.north west) {\bf Box 1: Transcriptome Age Index};
      \end{tikzpicture}
    \end{center}
  }

\date{}

\begin{document}

% Title must be 150 characters or less
\begin{flushleft}
{\Large
\textbf{The hourglass and the early conservation models - co-existing patterns of developmental constraints in vertebrates}
}
% Insert Author names, affiliations and corresponding author email.
\\
Barbara Piasecka$^{1,2,4}$, 
Pawe{\l} Lichocki$^{3}$, 
S\'{e}bastien Moretti$^{1,5}$,
Sven Bergmann$^{2,4,\#}$
Marc Robinson-Rechavi$^{1,4,\#,\ast}$
\\
\bf{1} Department of Ecology and Evolution, University of Lausanne, Lausanne, Switzerland
\\
\bf{2} Department of Medical Genetics, University of Lausanne, Lausanne, Switzerland
\\
\bf{3} Laboratory of Intelligent Systems, EPFL, Lausanne, Switzerland
\\
\bf{4} Swiss Institute of Bioinformatics, Lasuanne, Switzerland
\\
\bf{5} Vital-IT, Swiss Institute of Bioinformatics, Lausanne, Switzerland
\\
\# These authors contributed equally to this work.\\
$\ast$ E-mail: marc.robinson-rechavi@unil.ch
\end{flushleft}

% Please keep the abstract between 250 and 300 words
\section*{Abstract}
Developmental constraints have been postulated to limit the space of feasible phenotypes and thus shape animal evolution. These constraints have been suggested to be the strongest during either early or mid-embryogenesis, which corresponds to the early conservation model or the hourglass model, respectively. Conflicting results have been reported, but in recent studies of animal transcriptomes the hourglass model has been favored. Studies usually report descriptive statistics calculated for all genes over all developmental time points. This introduces dependencies between the sets of compared genes, and may lead to biased results. Here we overcome this problem using an alternative modular analysis. We used the Iterative Signature Algorithm to identify distinct modules of genes co-expressed specifically in consecutive stages of zebrafish development. We then performed a detailed comparison of several gene properties between modules, allowing for a less biased and more powerful analysis. Notably, our analysis corroborated the hourglass pattern only at the regulatory level, with sequences of regulatory regions being most conserved for genes expressed in mid-development, but not at the level of gene sequence, age or expression, in contrast to some previous studies. The early conservation model was supported with gene duplication and birth that were the most rare for genes expressed in early development. Finally, for all gene properties we observed the least conservation for genes expressed in late development or adult, consistent with both models. Overall, with the modular approach, we showed that different levels of molecular evolution follow different patterns of developmental constraints. Thus both models are valid, but with respect to different genomic features.

% Please keep the Author Summary between 150 and 200 words
% Use first person. PLoS ONE authors please skip this step. 
% Author Summary not valid for PLoS ONE submissions.   
\section*{Author Summary}
During development, vertebrate embryos pass through a ``phylotypic'' stage, during which their morphology is most similar between different species. This gave rise to the hourglass model, which predicts the highest developmental constraints during mid-embryogenesis. In the last decade a large effort has been made to uncover the relation between developmental constraints and the evolution of genome. Several studies reported gene characteristics that change according to the hourglass model, e.g., sequence conservation, age or expression. Here, we first show that some of the previous conclusions do not hold out under detailed analysis of the data. Then, we discuss the disadvantages of the standard evo-devo approach, i.e., comparing descriptive statistics of all genes across development. Results of such analysis are biased by genes expressed constantly during development (housekeeping genes). To overcome this limitation we use a modularization approach, which reduces the complexity of the data and assures independency between the sets of genes which are compared. We identified distinct sets of genes (modules) with time-specific expression in zebrafish development and analyzed their conservation of sequence, gene expression and regulatory elements, as well as their age, and orthology relationships. Interestingly, we found different patterns of developmental constraints for different genes properties. Only conserved regulatory regions follow an hourglass pattern.

\section*{Introduction}
Developmental constraints have been suggested to play an important role in shaping the evolution of embryonic development in animals. Briefly, the concept of developmental constraints assumes that the scope of developmental mechanisms limits the set of phenotypes that may evolve. Thus, morphological similarities between embryos of different species could reflect these underlying constraints \cite{Poe2004}. Two main models of embryonic developmental constraints have been put forward. The \emph{early conservation} model predicts that the highest developmental constraints occur at the beginning of embryogenesis. This corresponds to von Baer's third law \cite{vonBaer1828}, postulating that embryos of different species progressively diverge from one another during ontogeny. However, in modern times, the highest morphological similarity between embryos of different species was observed in the \emph{phylotypic stage} (i.e., mid-embryogenesis) \cite{Seidel1960, Sander1983, Elinson1987}. Consequently, Duboule \cite{Duboule1994} and Raff  \cite{Raff1996} proposed the so-called \emph{hourglass} model, which has since become widely accepted (see, e.g., \cite{Prudhomme2010, Kalinka2012}). It predicts the highest developmental constraints during mid-embryogenesis.

At the genomic level, the hourglass model was originally linked to the expression of HOX genes in animals \cite{Duboule1994}. More recently, the emphasis has shifted to the relation, if any, between developmental constraints and the evolution and function of the genome (reviewed in \cite{Kalinka2012}). Different studies have reported several characteristics supporting the hourglass model in animals on the genomic level. Hazkani-Covo et al. \cite{Hazkani-Covo2005} reported the highest protein sequence similarity between mouse and human for genes expressed in mid-development. In two influential papers, Domazet-Lo\v{s}o and Tautz \cite{Domazet-Loso2010} reported that the genes expressed in mid-development of zebrafish are older than genes expressed early or late, while Kalinka et al. \cite{Kalinka2010} showed that genes expressed in mid-development of fruit flies have the highest expression conservation. Similarly,  Irie and Kuratani \cite{Irie2011} reported the highest expression conservation between zebrafish, frog, chicken and mouse, for genes expressed in mid-development. Very recently, the hourglass model was argued to hold also for plants embryogenesis with respect to gene age and sequence conservation \cite{Quint2012}. However, some of these results do not hold out under a detailed analyses (see Box 1 and Supplementary Materials). For example, applying a standard log-transformation \cite{McDonald2009, Speed2000} to microarray signal intensities used in \cite{Domazet-Loso2010} changes the reported pattern such that it no longer supports the hourglass model (figure 1). Moreover, other studies have also found genetic patterns supporting an early conservation model \cite{Roux2008, Comte2010}.  

In most of the studies of developmental constraints the authors compared descriptive statistics of all genes across all developmental time-points (e.g., median expression \cite{Roux2008}, weighted mean age \cite{Domazet-Loso2010}, mean expression correlation \cite{Irie2011}). Such an approach introduces dependencies between the sets of genes which are compared, and consequently can produce results biased by genes expressed at many time-points. For example, housekeeping genes contribute to the average gene expression at all time points, and hence dilute trends. To overcome this essential problem, we have used a \emph{modularization} approach, which we applied to the recently published transcriptome data of zebrafish development \cite{Domazet-Loso2010}. We decomposed the genes into independent sets, i.e., \emph{modules}, that contained genes overexpressed solely in one of seven developmental stages: cleavage/blastula, gastrula, segmentation, pharyngula, larva, juvenile and adult. This decomposition allowed us to compare only sets of genes that have specific functions during embryonic development. For each of the seven modules, we studied five properties of its genes: 1) gene sequence conservation, 2) gene age, 3) gene expression conservation, 4) gene orthology relationships, and 5) regulatory elements conservation. 

Here, we show that different levels of molecular evolution follow different patterns of developmental constraints. First, the regulatory elements are most conserved for transcription factors expressed at mid-development, consistent with the hourglass model. Contrary to what has been reported previously \cite{Hazkani-Covo2005,Domazet-Loso2010, Irie2011}, we did not detect the hourglass pattern for gene sequence, age and expression. Second, constraints on gene duplication and on new gene introduction are the strongest in early development, supporting the early conservation model (consistent with \cite{Roux2008}). Finally, all gene properties displayed the least conservation in late development and adult, which is in agreement with both models of developmental constraints. 

% Results and Discussion can be combined.
\section*{Results}
\subsection*{Modules}
Our goal was to analyze the developmental constraints acting on different gene properties. To this end we identified and analyzed groups of genes co-expressed during distinct developmental stages. We applied the Iterative Signature Algorithm (ISA) \cite{Bergmann2003,Ihmels2004} to the zebrafish expression data published by Domazet-Lo\v{s}o and Tautz \cite{Domazet-Loso2010}, which measured the dynamics of the transcriptome during development with a resolution of 60 time points. The ISA is a modularization algorithm that finds genes with similar expression profiles and groups them into so-called transcription modules. In order to detect modules of genes with specific expression during the zebrafish development, we initialized the ISA with seven idealized expression profiles that corresponded to successive developmental stages (see Supplementary Materials and Supplementary figure S10).

We obtained seven modules, each containing genes overexpressed during one of the following developmental stages: cleavage/blastula, gastrula, segmentation, pharyngula, larva, juvenile and adult (figure \ref{fig:0}). Overall, the modules covered the entire development. The phylotypic stage in which the hourglass model predicts the highest evolutionary constraint corresponds to the segmentation and pharyngula modules. We will refer to these two modules as phylotypic modules. The cleavage/blastula and gastrula modules will be referred to as early modules, and larva, juvenile and adult modules as late modules.

The adjacent modules partially overlapped in their gene content. In order to allow for unbiased cross-module comparisons, genes belonging to two modules were kept in the one with the highest ISA gene score (see Methods); this concerned $534$ genes in total. The seven modules, i.e., cleavage/blastula, gastrula, pharyngula, segmention, larva, juvenile and adult, contained $444$, $820$, $487$, $414$, $415$, $290$ and $207$ genes, respectively (see Supplementary table S3 for the lists of the genes). Overall, $3077$ different genes were present in these modules, which implies a significant reduction of the number of genes being analyzed in comparison to the original data ($14 293$ genes on the microarray). In particular, the ISA removed the bias related to the genes expressed uniformly across development (like housekeeping genes). 

\subsection*{Functional annotation}
We verified the function of genes in modules detected by the ISA by comparing them to relevant known lists of genes. We found that the cleavage/blastula module was significantly enriched in maternal genes identified in \cite{Aanes2011} (36 genes vs. 19 expected by chance; hypergeometric test, $p = 0.01$), and the gastrula module was highly significantly enriched in post-midblastula transition (post-MBT) genes identified in \cite{Aanes2011} (78 genes vs. 25 expected by chance; hypergeometric test, $p = 2.8\times10^{-18}$). We confirmed the relevance of the segmentation and pharyngula modules by verifying that they were enriched in HOX genes (24 and 7 genes vs. 1 expected by chance, respectively; hypergeometric test, $p = 5.6\times10^{-16}$ and $2.9\times10^{-4}$, respectively), which is consistent with their role in mid-development \cite{Krumlauf1994}. We did not have any gold standard for genes expressed at the late stages of development.  However, since the early and phylotypic modules were enriched in genes with relevant functions, we are confident that the same is true for the late modules. 

Moreover, GO enrichment analysis confirmed that genes from the modules were enriched in functions relevant to the respective developmental stages. For example, the cleavage/blastula module was enriched in genes involved in protein phosphorylation and dephosphorylation processes, which is consistent with kinase-dependent control of cell cycle and regulation of mid-blastula transition (MBT) in vertebrates \cite{Hartley1996, Yarden1996}. The pharyngula module was enriched in genes associated with cell differentiation, and anatomical structure development. Finally, the adult module was enriched in genes involved in responses to environment, although not significantly (Supplementary table S2). 

\subsection*{Sequence conservation}
We checked whether the sequences of genes from different modules evolved under different selective pressure. To this end, we calculated the non-synonymous to synonymous substitution ratios ($d_N/d_S$) for genes in the modules and asked if the ratio was significantly lower for any of them. With the early conservation model, we would expect the lowest $d_N/d_S$ values for genes from early modules. Whereas with the hourglass model, we would expect the lowest $d_N/d_S$ values for genes from the phylotypic modules. 
In the cleavage/blastula module the median $d_N/d_S$ was not different from the median $d_N/d_S$ for all genes (equal to 0.15). In the other four modules covering embryonic development  the median $d_N/d_S$ was lower than the median $d_N/d_S$ for all genes (figure \ref{fig:1}A), and the difference was significant for all but the segmentation module (randomization test, $p < 0.003$ for the gastrula, pharyngula and larva modules). In the juvenile module, the median $d_N/d_S$ was significantly higher than the median $d_N/d_S$ for all genes (randomization test, $ p = 0.003$). In the adult module, the median $d_N/d_S$ was also higher than the median $d_N/d_S$ for all genes, but the difference was not significant. When analyzing separately sites under purifying selection or evolving neutrally, we also find weaker purifying selection during post-embryonic stages (see Supplementary Materials and Supplementary figure S11).

These results were consistent with the study by Roux and Robinson-Rechavi \cite{Roux2008}, who also reported equally low $d_N/d_S$ values during the entire zebrafish embryogenesis, and a small increase in mid-larva, juvenile and adult. In contrast, Hazkani-Covo et al. \cite{Hazkani-Covo2005} reported an hourglass pattern for protein distance between mouse and human genes expressed during development. However, the trend was not significant. In \cite{Roux2008} some evidence for early conservation was reported in mouse. Projecting the genes from zebrafish modules to mouse-human orthologs, we found equal conservation across development (Supplementary figure S12). Overall, data analyses support similar evolutionary constraints on sequences of genes expressed during whole embryogenesis of zebrafish, while for mouse more developmental data is needed to be conclusive.

\subsection*{Gene age}
The differences in age of genes expressed during different stages of the development have been suggested to be a good indicator of evolutionary constraints \cite{Irie2007,Domazet-Loso2010}. Thus, we investigated the age of genes belonging to different modules. We dated each gene by its first appearance in the phylogeny and assigned it to one of the five age groups: 1) Fungi/Metazoa, 2) Bilateria, 3) Coelomata+Chordata, 4) Euteleostomi and 5) Clupeocephala+\emph{Danio rerio}. Next, for each module we calculated the age distribution of its genes, i.e., the number of genes belonging to each age group, and compared it with the age distribution of all genes. 

For all but the cleavage/blastula module we detected significant age variations which differed across modules (figure \ref{fig:1}B; chi-square goodness of fit test, all $p < 1.3\times10^{-5}$). The oldest genes which belong to the Fungi/Metazoa class were overrepresented in the gastrula module (36.7\% of genes in the module vs. 25.7\% of all genes). The younger Bilateria genes were overrepresented in the phylotypic modules (45.5\% and 52.1\% of genes in the segmentation and pharyngula modules, respectively, vs. 34.4\% of all genes). The youngest genes were overrepresented in the late modules (e.g., for Eutelostomi genes: 25.7\%, 35.1\% and 35.6\% of such genes in larva, juvenile and adult modules, respectively, vs. 18\% of all genes). In contrast, Domazet-Lo\v{s}o and Tautz \cite{Domazet-Loso2010} reported that genes expressed in early and late development tend to be younger than genes expressed in mid-development, supporting the hourglass model. Yet, that result does not hold for log-transformed gene expression levels (Box 1), and is not recovered with measures of gene age other than the transcriptome age index (see Supplementary Materials and Supplementary figure S6). With the modular approach we observed that the age of expressed genes decreased throughout ontogeny. This pattern suggests that the oldest evolutionary stages tend to express the oldest genes.

\subsection*{Gene family size} 
Both gene duplication and gene loss can impact phenotypic evolution \cite{Ohno1970, Zhang2003, Nei2007, Wang2006, Demuth2009}. The outcome of these events can be summarized by the resulting gene family size. Consequently, constrained developmental stages should display less changes in gene family size than other stages. To test this hypothesis, for each zebrafish module we calculated  the number of its genes that were in 1) one-to-one, 2) one-to-many, 3) many-to-many, and 4) no orthology relation to mouse genes (i.e., no ortholog detectable by the criteria used in Ensembl Compara \cite{Vilella2009}). 

We compared the observed distributions with the distribution of the ortholog relationships for all genes. We detected significant variations of the ortholog relationship for the cleavage/blastula module and for all three late modules (chi-square goodness of fit test, all $p < 9\times10^{-5}$). Moreover, the pattern of variation itself differed across different modules. The number of one-to-one orthologs decreased throughout development (figure \ref{fig:1}C). It was significantly higher than expected only in the cleavage/blastula module ($54.6\%$ of genes in the module vs. $45.4\%$ of all genes). In contrast, the number of genes with no orthologous relationship increased throughout development (figure \ref{fig:1}C). It was significantly higher than expected only in the juvenile and adult modules (38.2\% and 38.4\% of genes in the two modules, respectively, vs. 20.4\% of all genes), consistent with the excess of ``young" genes. A similar pattern was observed for many-to-many orthologs (10.4\% and 7.8\% of genes in the two modules, respectively, vs. 3.9\% of all genes). Finally, the number of one-to-many orthologs was higher than expected only in the larva module (45.6\% of genes in the module vs. 30.3\% of all genes), and did not differ from expectation in all other modules. 

These results were consistent with \cite{Roux2008} in which the genes retained in duplicates after the teleost-specific whole genome duplication were reported to have low expression early in the development. Here, we recovered an analogous pattern with the modular approach, showing that the genes expressed early in the development are retained in duplicates less often than genes expressed later. Note that our observation is not limited to whole genome duplication. In addition, we detected the highest number of novel genes amongst genes expressed late in the development. 

\subsection*{Expression conservation}
Changes in gene expression are one of the main sources of morphological variation \cite{King1975,Preuss2004, Carroll2005}. The developmental constraints on gene expression might differ from those on the gene sequence \cite{Jordan2004, Yanai2004, Jordan2005}. Thus, for each module, we compared the mean expression profile of its genes with the mean expression profile of their one-to-one orthologs in mouse. We used two different data sets \cite{Wang2004, Irie2011} with expression values of mouse genes during the development. The use of two data sets was necessary, because there does not exist a single experiment covering the entire mouse development. The incompatibility of the two microarrays impaired the statistical strength of the analysis. For this reasons the results reported here should be regarded rather as qualitative than quantitative.

Since homology cannot be defined for individual developmental stages between zebrafish and mouse, we first mapped every time point to its broad metastage defined in Bgee \cite{bastian2008bgee} (figure \ref{fig:2}). Next, we calculated the mean expression level in every metastage. This resulted in six expression values for each gene during the development of mouse and zebrafish: zygote, cleavage, blastula, neurula, organogenesis, and post-embryonic stage. Note that the mouse microarrays did not cover the gastrula stage at all. For each module we calculated the Pearson's correlation between the mean expression of its genes and their mouse orthologs across the six metastages. For the cleavage/blastula module no correlation was detected, probably due to the incompatibility of the two mouse microarrays. Nevertheless, there exists a plausible, biological interpretation of the differences in gene expression between the early stages of zebrafish and mouse development. Zebrafish and mouse form two different embryological structures during blastulation, a blastula and a blastocyst, respectively. The blastocyst is a mammalian innovation that consists of an embryoblast (that develop into structures of the fetus) and a trophoblast (that form the extraembryonic tissue). In contrast, there is no extraembryonic tissue in zebrafish. Overall, the lack of correlation between gene expression for the early stages of mouse and zebrafish development could be explained by these structural differences.
For other modules the correlation was positive (figure \ref{fig:1}D), however due to the low number of data points in the analysis, no correlation values were significant (all $p>0.01$).

These results stood in contrast with the report by Irie and Kuratani \cite{Irie2011} who showed the highest conservation of gene expression in mid-development. However, a re-analysis of their data suggested that this observation was not significant (see Supplementary Materials and Supplementary figure S9). Also, both their and our studies shared problems related to the use of two data sets from different sources to cover mouse development. This and the lack of a straightforward homology between ontogenies of different species made it difficult to conclude on the conservation of gene expression during vertebrate development.

\subsection*{Regulatory regions}
The \emph{cis}-regulatory hypothesis asserts that most morphological evolution is due to changes in \emph{cis}-regulatory sequences \cite{Stern2000, Wray2007, Carroll2008}. A reasonable prediction of this hypothesis is slower \emph{cis}-element turnover in morphologically conserved developmental periods. We examined the presence of highly conserved non-coding elements (HCNEs) \cite{Engstrom2008} and of transposon-free regions (TFRs) \cite{Simons2007} in the proximity of genes from each module. In the analysis of HCNEs, we counted their number between zebrafish and mouse (detected with $70\%$ identity) in regions of $500$ base pairs upstream from the transcription start site. We found that only genes from the phylotypic modules were significantly enriched in HCNEs (hypergeometric test, $p = 8\times10^{-6}$, and $p = 1.1\times10^{-4}$ for segmentation and phayrngula modules, respectively). We tested the sensitivity of the results by changing the analyzed regions' length to $200$ and $1000$ base pairs upstream from the transcription start site, by looking for HCNEs in introns, and using HCNEs detected with identity of $90\%$. In all cases, we obtained similar results (see Supplementary table S1). In the analysis of TFRs, we counted the number of genes from each module that have been associated with TFRs in zebrafish. Importantly, these TFRs were reported to be conserved between vertebrates as distant as zebrafish and human. We found that only genes from the pharyngula module were significantly enriched in TFRs (hypergeometric test, $p = 5.7\times10^{-7}$).

The highly conserved non-coding elements and transposon-free regions are often associated with developmental regulatory genes, and with transcription factors (TFs) in particular \cite{Sandelin2004, Woolfe2005, Vavouri2007, Engstrom2008, Simons2007}. In order to confirm this association, we calculated the fractions of genes with HCNEs or with TFRs in their proximity. We observed that for both features this fraction was higher for TFs than for all genes. Importantly, we observed that only the phylotypic modules were enriched in TFs (figure \ref{fig:1}E). This partially explained the enrichment in HCNEs and TFRs for genes expressed in mid-development. In addition, HCNEs were more often present in the proximity of TFs from the pharyngula module than in the proximity of TFs in general (figure \ref{fig:1}E; $8.8\%$ of TFs from the pharyngula module had at least one HCNE in their proximity, and only $3.7\%$ of all TFs had at least one HCNEs in their proximity). Also TFRs were more often present in the proximity of TFs from the phylotypic modules than in the proximity of TFs in general (figure \ref{fig:1}E; $31\%$ and $45\%$ of TFs from the segmentation and pharyngula modules, respectively, had TFRs in their proximity, and only $26\%$ of all TFs had TFRs in their proximity).  Consequently, the enrichment in HCNEs and TFRs for genes expressed in the phylotypic stage seems to be related to the regulation of developmental processes. Interestingly, only few Hox genes from phylotypic modules were associated with HCNEs (four Hox genes from segmentation module), and with TFRs (six Hox genes from segmentation module, and one Hox gene from pharyngula module).

In addition, we checked for genes that preserved their specific ancestral order in the genome across metazoans (so called conserved ancestral microsyntenic pairs, \cite{Irimia2012}) and are known to be involved in the regulation of development. We found that they were slightly overrepresented in the segmentation module, but only at the limit of statistical significance (see Supplementary Materials).

Finally, we checked for core developmental genes in each module (see \cite{Vavouri2007} for the  list of genes). These genes are known to be involved in the regulation of development, and to have highly conserved regulatory regions within different taxa, including, nematodes, insects and vertebrates \cite{Vavouri2007}. We detected a significant enrichment in these genes only in the pharyngula module ($20$ core genes; hypergeometric test, $p = 6.9 \times 10^{-19}$), supporting the hourglass model.

\section*{Discussion}
Our goal was to study developmental constraints acting on various gene properties in vertebrates. Overall, we analyzed and compared five gene characteristics, namely the conservation of gene sequence, gene expression, and regulatory elements, as well as age and orthology relationships. To this end we identified distinct sets of genes with time-specific expression in zebrafish development, i.e., genes over-expressed in one of the seven consecutive stages: cleavage/blastula, gastrula, segmentation, pharyngula, larva, juvenile and adult. 
We believe that the change in expression level is a reliable indicator of gene involvement in different stages, although genes might also play a role outside the stages of their highest expression. Moreover, the modules contained genes overexpressed in relation to other stages, regardless of the absolute values of their expression. Thus, lowly expressed genes were also considered by the modularization algorithm, as long as they displayed some variance in expression levels over developmental time.
  
Several features do not show any significant pattern over embryonic development, often in contradiction to previous reports. There is notably no evidence for change in selective pressure acting on sequences of protein-coding genes (i.e., $d_N/d_S$) over development (in contrast to \cite{Hazkani-Covo2005}). Unfortunately, the available data does not allow a strong conclusion concerning the conservation of expression (in contrast to \cite{Irie2011}), despite the probable importance of this feature in the evolution of development. In this respect, the situation in vertebrates stands in contrast to the relatively clear results in flies \cite{Kalinka2010}, where the evolution of expression has been shown to be most constrained in mid-development.

Gene orthology relations support the early conservation model. We show that early stages are less prone to tolerate both gene duplication (consistent with \cite{Roux2008}) and gene introduction. The deficit in duplication in early development could also be due to a lack of opportunities for neo- or sub-functionalization in the anatomically simpler stages, which is not exclusive with strong purifying selection. The interpretation of transcriptome age is less straightforward. Our observations suggest that the oldest evolutionary stages tend to express of the oldest genes. It is possible that early stages are evolutionarily oldest, and that this is why they are enriched in oldest genes. Consequently, it is the presence of young genes in a module that would mark relaxed developmental constraints during the corresponding stage. However, neither early nor phylotypic modules are enriched in young genes (Euteleostomi and Clupeocephala+\emph{Danio rerio}), which suggests similar developmental constraints in early and mid-ontogeny. In any case, we do not find any support for the hypothesis that the phylotypic stage would be characterized by the oldest transcriptome (in contrast to \cite{Domazet-Loso2010}).

While the modularization approach does not support several previous hypotheses of genomic traces of the phylotypic period, it allows us to distinguish a strong signal of conservation of gene regulation in mid-development. While this had not yet been reported in genomic studies, it is consistent with early descriptions of the phylotypic stage as characterized by HOX genes body patterning activity \cite{Duboule1994}. Of note, the patterns that we observe are robust to the removal of Hox genes (data not shown), so they are more general than this original observation. We observed an excess of HCNEs only for genes expressed in the pharyngula module, and an excess of TFRs only for genes expressed in the phylotypic modules. The enrichment in HCNEs and TFRs has been related to developmental regulatory genes, and to transcription factors (TFs) in particular \cite{Sandelin2004, Woolfe2005, Vavouri2007, Engstrom2008}. Indeed, we observed that more TFs were expressed in mid-development than in other stages.  Also, we showed that a significant proportion of TFs expressed in mid-development had conserved regulatory regions (i.e., HCNEs and TFRs), in contrast to TFs expressed early or late. Consequently, the enrichment in HCNEs and TFRs for genes expressed in mid-development can be explained by both a higher number of TFs and a higher number of HCNEs and TFRs for these TFs, than for genes expressed earlier or later. Moreover, the pharyngula module was associated with core developmental genes. Overall, these results suggest that mid-developmental processes have extremely high conservation of regulation. This conservation could translate into observed common traits of the phylum expressed at the phenotypic level during mid-development. In addition, core developmental genes are known to be present in different taxa (e.g., nematodes, insects and vertebrates), in each of which they have a conserved regulation that evolved in parallel \cite{Vavouri2007}. This could explain why the phylotypic stage is observed not only in vertebrates \cite{Kimmel1995}, but also in other phyla, e.g., in arthropods \cite{Sander1983, Kalinka2010}.

Finally, for all of the features which we have considered there is at least some trend towards weaker evolutionary constraints in the latest stages: $d_N/d_S$ is higher in post-embryonic stages and there are less sites under purifying selection (Supplementary figure S11); correlation of expression is lowest for maternal, larval and adult genes; young genes and genes with duplications in fishes or other vertebrates are overrepresented in late modules; and genes expressed in juveniles and adults have the less HCNEs and TFRs. Although not all of these trends are significant, no feature shows stronger conservation in late development or adult. Thus, while different aspects of gene evolution show constraints at different times of development, there appears to be a generally faster evolution of all aspects of larval, juvenile and adult genes. Whether this is due to lower constraints (i.e., less purifying selection) or to stronger involvement in adaptation (i.e., more diversifying selection), remains an open question.

In summary, we studied evidence for, or against, any particular pattern of developmental constraints by considering sets of genes with time-specific expression patterns. Comparing such independent sets of genes with a clear function during embryogenesis resulted in cleaner and more fine-grained characterization of evolutionary patterns than previously reported. Notably, we showed that different levels of molecular evolution follow different patterns of developmental constraints. The sequence of regulatory regions is most conserved for genes expressed in mid-development, consistent with the hourglass model. Gene duplication and new gene introduction is most constrained during early development, supporting the early conservation model. Whereas, all  gene properties coherently show the least conservation for the latest stages, consistent with both the early conservation and the hourglass models.
 
% You may title this section "Methods" or "Models". 
% "Models" is not a valid title for PLoS ONE authors. However, PLoS ONE
% authors may use "Analysis" 
\section*{Methods}
\subsection*{Gene expression data}
Microarray data of zebrafish development were downloaded from NCBI's Gene Expression Omnibus \cite{Edgar2002} (GSE24616). This study was performed on the Agilent Zebrafish (V2) Gene Expression Microarray. In total, expression profiles for 60 developmental stages (from unfertilized egg to adults stages) were measured. The last ten stages (55 days - 1 year 6 months) were measured separately for male and female. Two replicates were made per time point, resulting in $(50 + 2 \times 10) \times 2 = 140$ microarrays in total. For each microarray, values of gProccessedSignal were log10 transformed and normalized as follows. Separately for each replicate, we equalized the expression signals between microarrays using the spike-ins reference, to account for different amounts of RNA present throughout development. To this aim, we first quantile normalized the expression signal of all spike-ins from all microarrays. Then, for each spike-in level we took the median value of expression signal before and after quantile normalization. This resulted in $10$ pairs of expression signals (original signal vs. normalized signal). With linear interpolation between these points, we obtained a piecewise linear curve that defined a mapping from original to normalized expression signals, which we used to equalize the expression signals from all microarrays. This was done by projecting each expression signal onto the piecewise linear curve and calculating the corresponding normalized value. Finally, we quantile normalized the data within replicates and computed the mean value for each gene within replicates. Expression values measured separately for males and females were averaged for each time point. 

Microarray data of mouse development were downloaded from Array Express (E-MEXP-51 and E-MTAB-368). The E-MEXP-51 study was performed on (C57BL/6$\times$CBA)F1  mice using Affymetrix GeneChip Murine Genome U74Av2. In total, expression profiles for 10 early developmental stages (zygote, early 2-cell, mid 2-cell, late 2-cell, 4 cell, 8 cell, 16 cell, early blastocyst, mid-blastocyst, late blastocyst) were measured. 2-4 replicates were made per time point. The data were normalized using gcRMA package.

The E-MTAB-368 study was performed on C57BL/6 mice using Affymetrix GeneChip Mouse Genome 430 2.0. In total, expression profiles for 8 mid and late developmental stages (E7.5, E8.5, E9.5, E10.5, E12.5, E14.5, E16.5, E18.5) were measured. 2-3 replicates were made per time point. The data were normalized using gcRMA package.

\subsection*{Mapping probe sets to Ensembl genes}
Agilent probe sets were mapped to their corresponding zebrafish genes (Ensembl release 63 \cite{Hubbard2009}) using BioMart \cite{Smedley2009}. Probe sets which did not map unambiguously to an Ensembl gene were excluded from the analysis. A total of 19 049 probe sets corresponding to 14 293 zebrafish genes were taken into account in our analysis.

Affymetrix probe sets were mapped to their corresponding mouse genes (Ensembl release 63 \cite{Hubbard2009}) using BioMart \cite{Smedley2009}. Probe sets which did not map unambiguously to an Ensembl gene were excluded from the analysis. For genes that were mapped by several probe sets we used the signal averaged across the probe sets. A total of 2883 mouse genes mapped by probe sets present on both mouse microarrays were taken into account in the gene expression analysis.

\subsection*{Iterative Signature Algorithm (ISA)}
The ISA identifies modules by an iterative procedure. A detailed description of the algorithm in the general case is given in \cite{Bergmann2003} (see also \url{http://www2.unil.ch/cbg/homepage/downloads/ISA_tutorial.pdf}). In this specific study, the algorithm was initialized with seven candidate seeds, each consisting of one artificial expression profile corresponding to one of the zebrafish developmental stages (see Supplementary Materials for details). Next, these seeds were refined through iterations by adding or removing genes and developmental time points until the processes converge to stable sets, which are referred to as (transcription) modules. Each developmental time point and gene received a score indicating their membership (if non-zero) and contribution to a given module. The closest the score for a gene or developmental time point was to one, the stronger the association between the gene/developmental time point and the rest of the module. 

The ISA was run twice with the following sets of thresholds: 1) $t_g = 1.8$ and $t_c = 1.2$, and 2) $t_g = 1.8$ and $t_c = 1.4$, for genes and developmental time points, respectively. We obtained the pharyngula module only in the case of $t_c=1.2$, and all other modules with both $t_c=1.2$ and $t_c=1.4$. All the modules contained their corresponding idealized profile. For further analysis, we kept a single module per developmental stage. From the pair of modules, we chose the one in which the idealized profile had a higher gene score. Overall, segmentation, pharyngula and juvenile modules were obtained with $t_c = 1.2$, and cleavage/blastula, gastrula, larva, and adult modules were obtained with $t_c = 1.4$.

\subsection*{GO enrichment analysis}
Gene ontology (GO) association for all genes mapped by zebrafish probe sets were downloaded from Ensembl release 63 \cite{Hubbard2009}, using BioMart \cite{Smedley2009}. GO enrichment was tested by Fisher's exact test, using the Bioconductor package topGO \cite{Alexa2006} version 2.2.0. The reference set consisted of all Ensembl genes mapped by probe sets of the microarray used. The ``elim'' algorithm of topGO was used to eliminate the (tree-like) hierarchical dependency of the GO terms. To correct for multiple testing the Bonferroni correction was applied. For every module GO categories with corrected P-value lower than $0.01$ were reported, if less then ten GO categories were significant we reported the top ten (see Supplementary table S2).

\subsection*{Gene sequence analysis} 
Ensembl Perl API release 70 \cite{Flicek2013} was used to extract all Ensembl Compara gene trees (and alignments) with a Clupeocephala (bony fishes) root. Sequences with too many gaps, or undefined nucleotides, were removed from the tree and alignment by MaxAlign (version 1.1) \cite{Gouveia-Oliveira2007}. Only trees without duplication (one-to-one orthologs) and with at least six leaves were kept. This resulted in 6769 trees. 

The site model from codeml \cite{Yang2007} (PAML package release 4.6; models M1a and M2a in codeml) was used to predict sites-specific selection in these trees. Finally, 916 trees were removed due to the lack of zebrafish genes, and 81 were removed due to lack of expression data on the zebrafish microarray. This resulted in 5772 trees. For every gene tree we calculated its mean $d_N/d_S$ value ($= p_0\omega_0+p_1\omega_1+p_2\omega_2$).

For every module we calculated the median $d_N/d_S$ ratio of its $k$ genes, where $k$ was the number of genes belonging to one of the 5772 trees. Next, we generated 10 000 sets of $k$ randomly chosen genes. For each set we calculated the median $d_N/d_S$ ratio. Thus, we constructed a sampling distribution of the median $d_N/d_S$ values for a set of $k$ genes. Then we calculated the probability that the median $d_N/d_S$ of the original module was sampled from the constructed distribution. This allowed us to assess if the observed median $d_N/d_S$ ratio was significantly different from the expected median value. To correct for multiple testing we applied the Bonferroni correction. We used $0.01$ as a significance level. 

\subsection*{Gene age analysis}
To study the age of genes belonging to different modules we dated the genes by their first appearance in the phylogeny. This consisted of retrieving the age of the oldest node of their Gene tree in Ensembl release 63 \cite{Hubbard2009}. Genes' age was described with one of the following categories: Fungi/Metazoa, Bilateria, Coelomata, Chordata, Eutelostomi, Clupeocephala, and \emph{Danio rerio}. To fit the chi-square test requirements (more than 5 elements in a group) we merged the genes into five age categories: Fungi/Metazoa, Bilateria, Coelomata + Chordata, Eutelostomi, Clupeocephala + \emph{Danio rerio}. Next, for every module we calculated the age distribution of its genes. We performed chi-square goodness of fit test to compare the observed and expected distributions of age classes in the modules. The expected distribution was estimated by classifying all zebrafish genes into one of the five age categories. To correct for multiple testing we applied the Bonferroni correction. We used $0.01$ as a significance level.

\subsection*{Zebrafish-Mouse orthologous genes}
Homology information of zebrafish and mouse genes was retrieved from Ensembl release 63 \cite{Hubbard2009}, using BioMart \cite{Smedley2009}. A total of 17 482 pairs of zebrafish-mouse orthologous genes had expression information in the zebrafish microarray data (14 293 zebrafish genes and 11 322 mouse genes). Among them there were 6441 one-to-one orthologous pairs, 5048 one-to-many orthologous pairs, and 2993 many-to-many orthologous pairs. 2901 zebrafish genes showed no orthology relationship with mouse genome. From further analysis we excluded 99 ``apparent-one-to-one" gene pairs. For every module we calculated the number of genes that were in one-to-one, one-to-many, many-to-many and no orthology relation to mouse genes. Next, we performed chi-square goodness of fit test to compare the observed and expected distributions of orthology classes in the modules. The expected distribution was estimated by classifying all zebrafish genes into one of the four orthology categories. To correct for multiple testing we applied the Bonferroni correction. We used $0.01$ as a significance level.

\subsection*{Gene expression conservation}
To study expression conservation between zebrafish genes assigned to the modules and their mouse one-to-one orthologs, we used gene expression data for 2883 orthologous gene pairs  (the limiting factor being the mapping to both mouse microarrays). For genes that were mapped by several probe sets we averaged their signal across the probe sets for both species. In order to compare gene expression between two species, we first calculated the mean expression for zebrafish genes present in the modules and their one-to-one mouse orthologs. Due to the incompatibility of two mouse microarray data used it was difficult to provide a meaningful comparison of expression for the two species. To calculate the correlation between expression profiles between zebrafish and mouse we reduced their expression profiles to six metastages: zygote, cleavage, blastula, neurula, organogenesis, and post-embryonic stage (see \cite{bastian2008bgee} for detailed definition of metastage). For every module and every metastage we calculated the mean expression level for zebrafish genes and their mouse one-to-one orthologs, and next we calculated the Pearson's correlation coefficient between them.  

\subsection*{Highly conserved non-coding elements}
Location data for highly conserved non-coding elements (HCNE) between zebrafish and mouse (70\% of identity) was retrieved from Ancora \cite{Engstrom2008} (\emph{http://ancora.genereg.net/downloads/danRer7/vs\_mouse}). The file \emph{HCNE\_danRer7\_mm9\_70pc\_50col.bed.gz} was downloaded and used in the analysis. For each of the 14 293 Ensembl genes considered in our analysis, we calculated the number of HCNE in regions of $500$ base pairs upstream from the transcription start site. Next, for every module we performed a hypergeometric test to assess if they were significantly enriched in genes with HCNE. To correct for multiple testing we applied the Bonferroni correction. We used $0.01$ as a significance level. In additional analyses, we calculated the number of HCNE in regions of $200$ and $1000$ base pairs upstream from the transcription start site, as well as in introns. Also, we repeated the analysis with HCNEs of $90\%$ identity (see Supplementary Materials).

\subsection*{Transposon-free regions}
Location data for transposon-free regions (TFRs) in zebrafish was retrieved from \cite{Simons2007} (\url{http://www.biomedcentral.com/content/supplementary/1471-2164-8-470-S1.txt}). First, each TFR was associated with Ensembl ID \cite{Hubbard2009} of its closest transcript from genome assembly Zv6. Then for each Ensembl transcript ID we retrieved an Ensembl gene ID from genome assembly Zv7. For every module we performed a hypergeometric test to asses if they were significantly enriched in genes with TFRs in their proximity. To correct for multiple testing we applied the Bonferroni correction. We used $0.01$ as a significance level.

\subsection*{Transcription factors}
The set of transcription factors was defined based on GO category annotation: GO: 0006355, regulation of transcription, DNA-dependent.
Among 14 293 Ensembl genes, 957 were annotated as transcription factors. For every module we performed a hypergeometric test to asses if they were significantly enriched in TFs. Next, we performed a hypergeometric test to asses if the TFs present in the modules were enriched in HCNEs and TFRs. To correct for multiple testing we applied the Bonferroni correction. We used $0.01$ as a significance level.

% Do NOT remove this, even if you are not including acknowledgments
\section*{Acknowledgments}
We thank Tim Hohm, Anna Kostikova, Zoltan Kutalik, Eyal Privman, and Pavan Ramdya and three anonymous reviewers for useful comments on the manuscript. We thank Gregory Barsh for helpful comments pre-review. We thank Julien Roux and all members of MRR and SB labs for helpful discussion. We thank Walid Gharib for help with $d_N/d_S$ computations. The computations were performed at the Vital-IT Center (\url{http://www.vital-it.ch}) for high-performance computing of the SIB Swiss Institute of Bioinformatics.

%\section*{References}

\boxabstract{
\bigskip 
Recent results of Domazet-Lo\v{s}o and Tautz \cite{Domazet-Loso2010} suggest that the oldest transcriptome set is expressed at the phylotypic stage, and that younger sets are expressed during early and late development, which support the hourglass model. To study the relationship between gene expression, ontogeny and phylogeny, the authors proposed a measure called the ``transcriptome age index'', or TAI. The TAI was defined as the mean of the phylogenetic ranks (``phylostrata'') across genes, weighted by their microarray signal intensity values at each developmental stage. Note that the microarray signal intensity values used in \cite{Domazet-Loso2010} displayed a log-normal distribution and spanned from $1$ to $10^5$ (Supplementary figure S1). Using these values to calculate TAI made the weights of phylogenetic ranks differ by five orders of magnitude between lowly and highly expressed genes. Consequently, only the most expressed genes (Supplementary figure S2), and potentially outliers (Supplementary figure S3), contributed to the hourglass pattern discovered with TAI. We found that applying a standard log-transformation to the intensity values changes the pattern, which then indicates older genes being expressed preferentially in early development (figure 1). The use of log-transformed data for microarray intensities is generally encouraged \cite{McDonald2009, Speed2000} because it keeps the biological signal, while removing dependency between variance and intensity of the analyzed signals. We present a more detailed re-analysis of the study of Domazet-Lo\v{s}o and Tautz \cite{Domazet-Loso2010} in Supplementary Materials (Supplementary figures S1 - S6). We also discuss in Supplementary Materials the study of Quint et al. \cite{Quint2012} that reported an hourglass pattern in plant embryogenesis using the same methodology (Supplementary figures S7 and S8).
\bigskip
\begin{center}
\emph{Here goes figure 1}
\end{center}
%{\bf Figure 1: Transcriptome age index (TAI) using raw and log-transformed expression signal intensities.} A higher TAI value implies that evolutionary younger genes are preferentially expressed at the corresponding time-point. The pink shaded area indicates the phylotypic stage. Colors of the curves reflect the main developmental periods and correspond to the colors used in \cite{Domazet-Loso2010}.
}
\setcounter{figure}{1}

\section*{Figure Legends}
{\bf Figure 1: Transcriptome age index (TAI) using raw and log-transformed expression signal intensities.} A higher TAI value implies that evolutionary younger genes are preferentially expressed at the corresponding time-point. The pink shaded area indicates the phylotypic stage. Colors of the curves reflect the main developmental periods and correspond to the colors used in \cite{Domazet-Loso2010}.

\begin{figure*}[!pht]
\caption{{\bf  Modules of genes with time-specific expression during zebrafish development.} A) Zebrafish ontogeny (drawings of the embryos are based upon sketches and photographs from \cite{Kimmel1995}. B) Median, 25th and 75th percentiles of expression value of genes in modules. Red bars denote the condition scores assigned to developmental points by the ISA.}
\label{fig:0}
\end{figure*}

\begin{figure*}[!pht]
\caption{{\bf  Measures of developmental constraints for various gene properties.} A) Box and Whisker plot showing non-synonymous to synonymous substitution ratios ($d_N/d_S$) for genes in the modules. The dotted line denotes median $d_N/d_S$ for all genes. The dash-dotted lines denote confidence interval for the median. B) Observed minus expected frequencies of age of genes in modules. C) Observed minus expected frequencies of orthology type (between zebrafish and mouse) for genes in modules. D) Mean expression level of zebrafish genes in modules, and their one-to-one orthologs in mouse in six developmental metastages. The transition between the two mouse data sets is denoted with the vertical dashed line. The Pearson's correlation coefficients for zebrafish and mouse expression profiles are reported for every module. E) The number of transcription factors (TFs) in modules (whole bar) and their enrichment in hihgly conserved non-coding elements (HCNEs) and transposon-free regions (TFRs). The stars denote significant enrichment ($p<0.01$) of TFs in HCNEs (yellow) and in TFRs (red). The dash-dotted lines denote confidence interval for the expected number of TFs in modules.}
\label{fig:1}
\end{figure*}

\begin{figure*}[!pht]
\begin{center}
\end{center}
\caption{{\bf Developmental metastages.} Mean expression level of zebrafish genes in modules, and of their one-to-one orthologs in mouse. The same colors denote corresponding developmental metastages in zebrafish and mouse.}
\label{fig:2}
\end{figure*}
\clearpage

\section*{Supplementary Figure Legends}
{\bf Figure S1. Total distribution of signal intensity from all $140$ microarrays \cite{Domazet-Loso2010}.}  \\
{\bf Figure S2. TAI hourglass pattern in zebrafish development \cite{Domazet-Loso2010} is driven by the subset of most highly expressed genes.} Removing the 20\% of top expressed genes at every developmental stage changes the overall pattern. Resulting TAI pattern has very low values and does not follow the hourglass shape any more (grey line). \\
{\bf Figure S3. Sensitivity to outliers.}  (A) Raw expression signal of probe A\_15\_P161596 across zebrafish development. (B) TAI calculated on non-transformed data across zebrafish development without this probe (red) and the effect of this probe on TAI pattern (grey). (C) TAI calculated on log10-transformed data across zebrafish development without this probe (red) and the effect of this probe on TAI pattern (grey). Expression data from \cite{Domazet-Loso2010}. \\
{\bf Figure S4. TAI calculated using expression intensities of genes, instead of probes, across zebrafish development.} For each gene we averaged the signal intensity from all corresponding probes. After this process 16 188 probes' intensities values were reduced to 12 892 genes' intensities values, which were used to weight the phylogenetic ranks of genes (if two different phylostrata were assigned to the same gene, the older one was chosen). (A) non-transformed data was used. (B) log10-transformed data was used. Expression data from \cite{Domazet-Loso2010}. \\
{\bf Figure S5. TAI calculated using genes recoded as present-absent across zebrafish development.} At a given stage of development, if the log10-intensity value of a gene is above one, its expression is set to 1, otherwise it is set to 0. Other notations as in figure 1 (in main text). Expression data from \cite{Domazet-Loso2010}. \\
{\bf Figure S6. Alternative measures of transcriptome age.}  (A) Mean age of genes expressed across zebrafish development; age estimated with the TimeTree database (\url{www.timetree.org}). A gene is considered expressed at a given stage of development if its log10-intensity is above one. (B) Difference between median expression profiles of old genes and young genes across zebrafish development. Here, the genes that have emerged before the evolution of Metazoa are considered old and the genes that have emerged since the ancestor of Euteleostomi are considered young. The difference between the two groups is always positive, reflecting that old genes tend to be more expressed than young genes \cite{Wolf2009}. The results are robust to the choice of cutoffs used to define old and young genes (data not shown). Red dashed line - female data, blue dashed line - male data. Other notations as in figure 1 (main text). Expression data from \cite{Domazet-Loso2010}. \\
{\bf Figure S7. TAI and TDI hourglass patterns in \emph{Arabidopsis} development \cite{Quint2012} are driven by a very small subset of the most highly expressed genes.} Removing only the 1\% of top expressed genes at each developmental stage changes the overall pattern. Resulting TAI and TDI patterns do not follow the hourglass shape any more (grey line). \\
{\bf Figure S8. TAI and TDI calculated using raw (green line) and log-transformed (grey line) expression signal intensities.} Data from \cite{Quint2012}. \\
{\bf Figure S9. Correlation between expression levels of genes across developmental time points of mouse, chicken and zebrafish}. Field A denotes the early stages, field B denotes the phylotypic stages, and field C denotes the late stages of development. Expression data from \cite{Irie2011}. \\
{\bf Figure S10. Artificial expression profiles used to initialize the ISA:} pre-MBT, post-MBT, ``middle'', pharyngula, larva, ``late'', adult. These profiles resulted in modules containing genes expressed specifically in: cleavage/blastula, gastrula, segmentation, pharyngula, larva, juvenile, and adult, respectively. \\
{\bf Figure S11. Measures of purifying selection for gene trees of bony fishes.} (A) Average dN/dS for sites under purifying selection ($\omega_0$). (B) Proportion of sites under purifying selection ($p_0$).\\
{\bf Figure S12. $d_N/d_S$ ratio for human-mouse one-to-one orthologs.} The orthologs were obtained by projecting the genes expressed in the zebrafish modules to their one-to-one orthologs in mouse and human.

\section*{Supplementary Table Legends}
{\bf Table S1. P-values from HCNE enrichment analyses.} \\
{\bf Table S2. The list of modules and their enriched GO categories (biological process).}\\
{\bf Table S3. The list of genes belonging to each module.}
%\section*{Tables}
%\begin{table}[!ht]
%\caption{
%\bf{Table title}}
%\begin{tabular}{|c|c|c|}
%table information
%\end{tabular}
%\begin{flushleft}Table caption
%\end{flushleft}
%\label{tab:label}
% \end{table}


\begin{thebibliography}{10}
\providecommand{\url}[1]{\texttt{#1}}
\providecommand{\urlprefix}{}
\expandafter\ifx\csname urlstyle\endcsname\relax
  \providecommand{\doi}[1]{doi:\discretionary{}{}{}#1}\else
  \providecommand{\doi}{doi:\discretionary{}{}{}\begingroup
  \urlstyle{rm}\Url}\fi
\providecommand{\bibAnnoteFile}[1]{%
  \IfFileExists{#1}{\begin{quotation}\noindent\textsc{Key:} #1\\
  \textsc{Annotation:}\ \input{#1}\end{quotation}}{}}
\providecommand{\bibAnnote}[2]{%
  \begin{quotation}\noindent\textsc{Key:} #1\\
  \textsc{Annotation:}\ #2\end{quotation}}
\providecommand{\eprint}[2][]{\url{#2}}

\bibitem{Poe2004}
Poe S, Wake MH (2004) Quantitative tests of general models for the evolution of
  development.
\newblock Am Nat 164: 415-22.
\input{Poe2004}

\bibitem{vonBaer1828}
von Baer KE (1828) Ueber Entwicklungsgeschichte der Thiere: Beobachtung und
  Reflexion.
\newblock K{\"o}nigsberg: Borntr{\"a}ger.
\input{vonBaer1828}

\bibitem{Seidel1960}
Seidel F (1960) K{\"o}rpergrundgestalt und keimstruktur. eine er{\"o}rterung
  {\"u}ber die grundlagen der vergleichenden und experimentellen embryologie
  und deren g{\"u}ltigkeit bei phylogenetischen {\"u}berlegungen.
\newblock Zool Anz 164: 245-305.
\input{Seidel1960}

\bibitem{Sander1983}
Sander K (1983) The evolution of patterning mechanisms: gleanings from insect
  embryogenesis and spermatogenesis.
\newblock In: Goodwin~BC WC Holder~N, editor, Development and evolution.
  Cambridge University Press, pp. 137-159.
\input{Sander1983}

\bibitem{Elinson1987}
Elinson R (1987) Change in developmental patterns: Embryos of amphibians with
  large eggs.
\newblock In: Raff~RA RE, editor, Development as an Evolutionary Process. New
  York: Alan R. Liss., pp. 1-21.
\input{Elinson1987}

\bibitem{Duboule1994}
Duboule D (1994) Temporal colinearity and the phylotypic progression: a basis
  for the stability of a vertebrate bauplan and the evolution of morphologies
  through heterochrony.
\newblock Dev Suppl : 135-42.
\input{Duboule1994}

\bibitem{Raff1996}
Raff RA (1996) The shape of life: genes, development, and the evolution of
  animal form.
\newblock Chicago; London: University of Chicago Press.
\input{Raff1996}

\bibitem{Prudhomme2010}
Prud'homme B, Gompel N (2010) Evolutionary biology: Genomic hourglass.
\newblock Nature 468: 768-9.
\input{Prudhomme2010}

\bibitem{Kalinka2012}
Kalinka A, Tomancak P (2012) The evolution of early animal embryos:
  conservation or divergence?
\newblock Trends in Ecology \& Evolution 27: 385-393.
\input{Kalinka2012}

\bibitem{Hazkani-Covo2005}
Hazkani-Covo E, Wool D, Graur D (2005) In search of the vertebrate phylotypic
  stage: a molecular examination of the developmental hourglass model and von
  {B}aer's third law.
\newblock J Exp Zool B Mol Dev Evol 304: 150-8.
\input{Hazkani-Covo2005}

\bibitem{Domazet-Loso2010}
Domazet-Lo{\v s}o T, Tautz D (2010) A phylogenetically based transcriptome age
  index mirrors ontogenetic divergence patterns.
\newblock Nature 468: 815-8.
\input{Domazet-Loso2010}

\bibitem{Kalinka2010}
Kalinka AT, Varga KM, Gerrard DT, Preibisch S, Corcoran DL, et~al. (2010) Gene
  expression divergence recapitulates the developmental hourglass model.
\newblock Nature 468: 811-4.
\input{Kalinka2010}

\bibitem{Irie2011}
Irie N, Kuratani S (2011) Comparative transcriptome analysis reveals vertebrate
  phylotypic period during organogenesis.
\newblock Nat Commun 2: 248.
\input{Irie2011}

\bibitem{Quint2012}
Quint M, Drost HG, Gabel A, Ullrich KK, B{\"o}nn M, et~al. (2012) A
  transcriptomic hourglass in plant embryogenesis.
\newblock Nature .
\input{Quint2012}

\bibitem{McDonald2009}
McDonald JH (2009) Handbook of Biological Statistics (2nd ed.).
\newblock Baltimore, Maryland: Sparky House Publishing.
\input{McDonald2009}

\bibitem{Speed2000}
Speed T (2000) Always log spot intensities and ratios.
\newblock
  \urlprefix\url{http://www.stat.berkeley.edu/users/terry/zarray/Html/log.html%
}.
\input{Speed2000}

\bibitem{Roux2008}
Roux J, Robinson-Rechavi M (2008) Developmental constraints on vertebrate
  genome evolution.
\newblock PLoS Genet 4: e1000311.
\input{Roux2008}

\bibitem{Comte2010}
Comte A, Roux J, Robinson-Rechavi M (2010) Molecular signaling in zebrafish
  development and the vertebrate phylotypic period.
\newblock Evolution \& development 12: 144--156.
\input{Comte2010}

\bibitem{Bergmann2003}
Bergmann S, Ihmels J, Barkai N (2003) Iterative signature algorithm for the
  analysis of large-scale gene expression data.
\newblock Phys Rev E Stat Nonlin Soft Matter Phys 67: 031902.
\input{Bergmann2003}

\bibitem{Ihmels2004}
Ihmels J, Bergmann S, Barkai N (2004) Defining transcription modules using
  large-scale gene expression data.
\newblock Bioinformatics 20: 1993-2003.
\input{Ihmels2004}

\bibitem{Aanes2011}
Aanes H, Winata CL, Lin CH, Chen JP, Srinivasan KG, et~al. (2011) Zebrafish
  m{RNA} sequencing deciphers novelties in transcriptome dynamics during
  maternal to zygotic transition.
\newblock Genome Res 21: 1328-38.
\input{Aanes2011}

\bibitem{Krumlauf1994}
Krumlauf R (1994) Hox genes in vertebrate development.
\newblock Cell 78: 191-201.
\input{Krumlauf1994}

\bibitem{Hartley1996}
Hartley RS, Rempel RE, Maller JL (1996) In vivo regulation of the early
  embryonic cell cycle in {X}enopus.
\newblock Dev Biol 173: 408-19.
\input{Hartley1996}

\bibitem{Yarden1996}
Yarden A, Geiger B (1996) Zebrafish cyclin {E} regulation during early
  embryogenesis.
\newblock Dev Dyn 206: 1-11.
\input{Yarden1996}

\bibitem{Irie2007}
Irie N, Sehara-Fujisawa A (2007) The vertebrate phylotypic stage and an early
  bilaterian-related stage in mouse embryogenesis defined by genomic
  information.
\newblock BMC Biol 5: 1.
\input{Irie2007}

\bibitem{Ohno1970}
Ohno S, et~al. (1970) Evolution by gene duplication.
\newblock Berlin, Heidelberg and New York: Springer-Verlag.
\input{Ohno1970}

\bibitem{Zhang2003}
Zhang J (2003) Evolution by gene duplication - an update.
\newblock Trends Ecol Evol 18: 292-298.
\input{Zhang2003}

\bibitem{Nei2007}
Nei M (2007) The new mutation theory of phenotypic evolution.
\newblock Proc Natl Acad Sci U S A 104: 12235-42.
\input{Nei2007}

\bibitem{Wang2006}
Wang X, Grus WE, Zhang J (2006) Gene losses during human origins.
\newblock PLoS Biol 4: e52.
\input{Wang2006}

\bibitem{Demuth2009}
Demuth JP, Hahn MW (2009) The life and death of gene families.
\newblock Bioessays 31: 29-39.
\input{Demuth2009}

\bibitem{Vilella2009}
Vilella AJ, Severin J, Ureta-Vidal A, Heng L, Durbin R, et~al. (2009)
  Ensembl{C}ompara genetrees: Complete, duplication-aware phylogenetic trees in
  vertebrates.
\newblock Genome Res 19: 327-35.
\input{Vilella2009}

\bibitem{King1975}
King MC, Wilson AC (1975) Evolution at two levels in humans and chimpanzees.
\newblock Science 188: 107-16.
\input{King1975}

\bibitem{Preuss2004}
Preuss TM, C{\'a}ceres M, Oldham MC, Geschwind DH (2004) Human brain evolution:
  insights from microarrays.
\newblock Nat Rev Genet 5: 850-60.
\input{Preuss2004}

\bibitem{Carroll2005}
Carroll SB (2005) Evolution at two levels: on genes and form.
\newblock PLoS Biol 3: e245.
\input{Carroll2005}

\bibitem{Jordan2004}
Jordan IK, Mari{\~n}o-Ram{\'\i}rez L, Wolf YI, Koonin EV (2004) Conservation
  and coevolution in the scale-free human gene coexpression network.
\newblock Mol Biol Evol 21: 2058-70.
\input{Jordan2004}

\bibitem{Yanai2004}
Yanai I, Graur D, Ophir R (2004) Incongruent expression profiles between human
  and mouse orthologous genes suggest widespread neutral evolution of
  transcription control.
\newblock OMICS 8: 15--24.
\input{Yanai2004}

\bibitem{Jordan2005}
Jordan IK, Marino-Ramirez L, Koonin EV (2005) Evolutionary significance of gene
  expression divergence.
\newblock Gene 345: 119--126.
\input{Jordan2005}

\bibitem{Wang2004}
Wang QT, Piotrowska K, Ciemerych MA, Milenkovic L, Scott MP, et~al. (2004) A
  genome-wide study of gene activity reveals developmental signaling pathways
  in the preimplantation mouse embryo.
\newblock Dev Cell 6: 133-44.
\input{Wang2004}

\bibitem{bastian2008bgee}
Bastian F, Parmentier G, Roux J, Moretti S, Laudet V, et~al. (2008) Bgee:
  Integrating and comparing heterogeneous transcriptome data among species.
\newblock In: Bairoch A, Cohen-Boulakia S, Froidevaux C, editors, Data
  Integration in the Life Sciences, Springer Berlin / Heidelberg, volume 5109
  of \emph{Lecture Notes in Computer Science}. pp. 124-131.
\input{bastian2008bgee}

\bibitem{Stern2000}
Stern DL (2000) Evolutionary developmental biology and the problem of
  variation.
\newblock Evolution 54: 1079-91.
\input{Stern2000}

\bibitem{Wray2007}
Wray GA (2007) The evolutionary significance of cis-regulatory mutations.
\newblock Nat Rev Genet 8: 206-16.
\input{Wray2007}

\bibitem{Carroll2008}
Carroll SB (2008) Evo-devo and an expanding evolutionary synthesis: a genetic
  theory of morphological evolution.
\newblock Cell 134: 25-36.
\input{Carroll2008}

\bibitem{Engstrom2008}
Engstr{\"o}m PG, Fredman D, Lenhard B (2008) Ancora: a web resource for
  exploring highly conserved noncoding elements and their association with
  developmental regulatory genes.
\newblock Genome Biol 9: R34.
\input{Engstrom2008}

\bibitem{Simons2007}
Simons C, Makunin IV, Pheasant M, Mattick JS (2007) Maintenance of
  transposon-free regions throughout vertebrate evolution.
\newblock BMC Genomics 8: 470.
\input{Simons2007}

\bibitem{Sandelin2004}
Sandelin A, Bailey P, Bruce S, Engstr{\"o}m PG, Klos JM, et~al. (2004) Arrays
  of ultraconserved non-coding regions span the loci of key developmental genes
  in vertebrate genomes.
\newblock BMC Genomics 5: 99.
\input{Sandelin2004}

\bibitem{Woolfe2005}
Woolfe A, Goodson M, Goode DK, Snell P, McEwen GK, et~al. (2005) Highly
  conserved non-coding sequences are associated with vertebrate development.
\newblock PLoS Biol 3: e7.
\input{Woolfe2005}

\bibitem{Vavouri2007}
Vavouri T, Walter K, Gilks WR, Lehner B, Elgar G (2007) Parallel evolution of
  conserved non-coding elements that target a common set of developmental
  regulatory genes from worms to humans.
\newblock Genome Biol 8: R15.
\input{Vavouri2007}

\bibitem{Irimia2012}
Irimia M, Tena JJ, Alexis M, Fernandez-Mi{\~n}an A, Maeso I, et~al. (2012)
  Extensive conservation of ancient microsynteny across metazoans due to
  cis-regulatory constraints.
\newblock Genome Res .
\input{Irimia2012}

\bibitem{Kimmel1995}
Kimmel CB, Ballard WW, Kimmel SR, Ullmann B, Schilling TF (1995) Stages of
  embryonic development of the zebrafish.
\newblock Dev Dyn 203: 253-310.
\input{Kimmel1995}

\bibitem{Edgar2002}
Edgar R, Domrachev M, Lash AE (2002) Gene {E}xpression {O}mnibus: {NCBI} gene
  expression and hybridization array data repository.
\newblock Nucleic Acids Res 30: 207-10.
\input{Edgar2002}

\bibitem{Hubbard2009}
Hubbard TJP, Aken BL, Ayling S, Ballester B, Beal K, et~al. (2009) Ensembl
  2009.
\newblock Nucleic Acids Res 37: D690-7.
\input{Hubbard2009}

\bibitem{Smedley2009}
Smedley D, Haider S, Ballester B, Holland R, London D, et~al. (2009) Bio{M}art
  -- biological queries made easy.
\newblock BMC Genomics 10: 22.
\input{Smedley2009}

\bibitem{Alexa2006}
Alexa A, Rahnenfuhrer J, Lengauer T (2006) Improved scoring of functional
  groups from gene expression data by decorrelating go graph structure.
\newblock Bioinformatics 22: 1600--1607.
\input{Alexa2006}

\bibitem{Flicek2013}
Flicek P, Ahmed I, Amode MR S, Barrell D, Beal K, et~al. (2013) Ensembl
  2013.
\newblock Nucleic Acids Res 41: D48--55.
\input{Flicek2013}

\bibitem{Gouveia-Oliveira2007}
 Gouveia-Oliveira R, Sackett PW, Pedersen AG (2007) Max{A}lign: maximizing usable data in alignment.
\newblock BMC Bioinformatics 8: 312.
\input{Gouveia-Oliveira2007}

\bibitem{Yang2007}
Yang Z (2007) P{AML} 4: Phylogenetic Analysis by Maximum Likelihood.
\newblock Mol Biol Evol 24: 1586--91.
\input{Yang2007}

\bibitem{Wolf2009}
Wolf YI, Novichkov PS, Karev GP, Koonin EV, Lipman DJ (2009) The universal
  distribution of evolutionary rates of genes and distinct characteristics of
  eukaryotic genes of different apparent ages.
\newblock Proc Natl Acad Sci U S A 106: 7273-80.
\input{Wolf2009}

\end{thebibliography}
\end{document}